\documentclass[aps,prb,twocolumn,groupedaddress,showpacs]{revtex4}
\usepackage[latin1]{inputenc}
\usepackage[english]{babel}
\usepackage{setspace}  
\usepackage{graphicx}
\usepackage{amssymb}
\usepackage{amsmath}
\usepackage{amscd}
\usepackage{rotating}
\usepackage{color}
\usepackage{epstopdf}

\begin{document}

\title{Distinguishing Phases with Ansatz Wavefunctions}

\author{B. Bauer}
\affiliation{
Theoretische Physik, ETH Zurich, 8093 Zurich, Switzerland
}
\author{V.W.  Scarola} 
\altaffiliation [Present Address:] {Physics Department, Virginia Tech, Blacksburg, Virginia 24061, USA}
\affiliation{
Theoretische Physik, ETH Zurich, 8093 Zurich, Switzerland
}
\affiliation{
Department of Chemistry and Pitzer Center for Theoretical Chemistry, University of California, Berkeley, California 94720, USA
}
\author{M. Troyer}
\affiliation{
Theoretische Physik, ETH Zurich, 8093 Zurich, Switzerland
}
\author{K.B. Whaley}
\affiliation{
Department of Chemistry and Pitzer Center for Theoretical Chemistry, University of California, Berkeley, California 94720, USA
}

\date{\today}

\begin{abstract}
We propose an indistinguishability measure for assessment of ansatz wavefunctions with numerically determined wavefunctions.  The measure efficiently compares all correlation functions of two states and can therefore be used to distinguish phases by defining
 correlator classes for ansatz wavefunctions.  It also allows identification of quantum critical points.  We demonstrate the approach for the transverse Ising and bilinear-biquadratic Heisenberg models, using the matrix product state formalism with the time evolving block decimation algorithm.
\end{abstract}


\maketitle


\section{Introduction}

A growing number of quantum many-body models have been constructed to study   
order that is readily characterized by wavefunctions as opposed to simple order parameters.  Models of topological order \cite{wen1990} in particular can be examined with a combination of numerical techniques and ansatz states.  
Examples include idealized models of the fractional quantum Hall regime \cite{haldane1983,trugman1985,jain2007} and spin liquids such as the 
AKLT model \cite{affleck1987} or the toric code \cite{kitaev2003}.  The exact ground states of these models, e.g., the Laughlin \cite{laughlin1983} or valence bond solid  \cite{affleck1987}  (VBS) wavefunctions,  
serve as ansatz states for more realistic many-body models.  
A comparison of idealized ansatz wavefunctions and numerically obtained, realistic wavefunctions  
then becomes an essential element in the search for novel phases in real materials.

Several procedures are currently available for verifying ansatz states.  The variational theorem
allows comparison of the energetics of proposed ground states.  It is useful in ruling out 
trial states but can fail in establishing a particular trial state because irrelevant 
wavefunctions often show competitive energetics.    
Diagonalization of small systems can be used to compute overlaps between ansatz and 
exact states.  But the scaling of overlap to larger systems does not always allow for a clear 
identification of a particular phase.  For example diagonalization studies \cite{morf1998} of proposed $\nu=5/2$  fractional 
quantum Hall states show overlaps of $\sim 0.8-0.9$ of competing paired states, complicating unambiguous identification of 
the true ground state.  Variational tests and 
overlaps must be combined with systematic analyses of 
other correlation functions to make a case for how well 
an ansatz state captures output from numerics.    
 
In the following, we present a new measure for assessment of ansatz wavefunctions by comparison with numerical wavefunctions that allows for a clear identification of a particular phase.  This measure, which we refer to as the indistinguishability, $I$, is based on a quantum information measure of quantum state distinguishability \cite{helstrom1976,fuchs1999,korsbakken2007}. We demonstrate the use of this indistinguishability measure by application to the assessment of ansatz wavefunctions for the ground state of two different models, the transverse Ising chain and the spin-one bilinear-biquadratic Heisenberg chain.  The former provides a simple test of the approach while the latter allows us to exploit its power to analyze a 
challenging and rich model whose solution has not yet been fully characterized.  
We use accurate ground state wavefunctions that are obtained with the time-evolving block decimation algorithm of Vidal \cite{vidal_TEBD,daley2004}. This yields the state in the form of a matrix-product state (MPS) \cite{ostromm,verstraete2008}, from which our measure can easily be calculated. 
We note that the notion of
quantum state distinguishability has been used recently to derive order parameters \cite{furukawa2005}.  We emphasize that we take an entirely different 
approach here by using distinguishability to assess the degree of similarity of a proposed ansatz wavefunction with output from an accurate simulation, thereby gaining insight from the structure of the wavefunctions.

We also show that the indistinguishability measure allows identification of quantum critical points and leads to an indistinguishability susceptibility that provides an accurate signature of these.  
Recent work has explored the characterization of quantum phase transitions without making recourse 
to ansatz wavefunctions.  Instead, quantities related to quantum information theory 
such as  concurrence \cite{osborne2002,osterloh2002}, entanglement entropy \cite{vidal2003} 
and fidelity \cite{zanardi2006,zhou2008} 
have been used to extract information about quantum phase transitions from numerical output.  These and other quantities signal changes of phase but without revealing detailed information about the nature of the quantum states. 

The outline of this paper is as follows: in Sect. \ref{sct:i} and \ref{sct:qcb}, we introduce the new measure and discuss the scaling in the thermodynamic limit. In Sect. \ref{sct:mps}, we show how to calculate the measure efficiently for a given state in the MPS representation. In Sect. \ref{sct:ising} and \ref{sct:blbq}, we discuss our results for the transverse Ising chain and the bilinear-biquadratic Heisenberg chain, respectively.


\section{Indistinguishability \label{sct:i}}

We define the indistinguishability $I_{n}(\text{A,E})$ of two $N$-particle states, an ansatz state $\Psi_\text{A}$ and the exact state $\Psi_\text{E}$, as the $n$-particle probability of 
error in distinguishing the two states with an $n$-particle measurement:
\begin{eqnarray}
I_{n}(A,E)=\frac{1}{2}-\frac{1}{4}\text{Tr} \vert \rho^{(n)}_{\text{E}}-\rho^{(n)}_{\text{A}} \vert
\label{PE}
\end{eqnarray} 
where $\text{Tr} \vert \mathcal{O} \vert$ is the trace norm of $\mathcal{O}$ and $\rho^{(n)}=\text{Tr}_{N-n}\left(\rho\right)$ is the $n$-particle 
reduced density matrix \cite{helstrom1976, fuchs1999,korsbakken2007}. 
The last term in Eq.~\ref{PE} is a well known statistical distance measure, the Kolmogorov distance 
between two probability distributions.  When 
$I_{n}(\text{A,E})$ is zero, the states are distinguishable and the ansatz wavefunction $\Psi_\text{A}$ is clearly a bad approximation to $\Psi_\text{E}$.  However, when it is non-zero, there is a finite probability that an $n$-particle measurement can not distinguish the ansatz from the numerical wavefunction, implying that 
the ansatz provides a good description of the state up to $n$-particle correlators.  $I_{n}(\text{A,E})=1/2$ corresponds to maximum indistinguishability, implying identical states.  Since the measure is defined in terms of reduced density matrices, the state indistinguishability implicitly
scans all correlators with up to $n$ particles, to yield a single number that quantifies the ability of an optimally chosen set 
of $n$-particle correlators to distinguish two states \cite{helstrom1976, fuchs1999}.  $1-I_{n}$ gives the probability that an optimally chosen correlation function involving at most $n$ particles will be able to distinguish the two states.  

We use $I_{n}$ as a quantifier of the degree of {\em indistinguishability} of two states via correlators in an $N$-particle system.  
When $I_{n}$ is intensive in $N$, so that small $n$ values suffice to characterize the correlators, we define 
two states to be in the same $n$-particle \emph{correlator class} if $I_{n}$ 
is finite in the thermodynamic limit, i.e., as $N \rightarrow \infty$, and to be in different correlator classes if $I_{n}=0$ in this limit.

\section{Quantum Chernoff bound \label{sct:qcb}}

It is possible that $I_n$ vanishes in the thermodynamic limit regardless of the ansatz. 
When $I_{n}$ is extensive in $N$ so that large $n$ values are required
(e.g., $n \sim \mathcal{O}(N)$), we 
can nevertheless use the scaling of $I_n$ with $N$ to identify correlator classes.  The scaling of the indistinguishability to the thermodynamic limit can be quantified in terms of the 
quantum Chernoff bound (QCB).  Assuming that on sufficiently large scales, a translationally invariant ground 
state can be regarded as a tensor product 
of subblocks (or copies), a recent result \cite{audenaert2007} for the indistinguishability of many copies
of the system shows that we should expect an
exponential dependence for large $n$, i.e.:  $I_n\sim \exp(-n\xi_{\text{CB}})$, where 
the QCB can be identified in the thermodynamic limit from
\begin{equation}
\xi_{\text{CB}}^{\text{lim}} = - \lim_{n \to \infty } \log(I_n)/n
\end{equation}
with $n=N/2$. 
A remarkable relation \cite{audenaert2007} connects the QCB directly to the reduced density matrices of \emph{finite} blocks, namely
\begin{equation}
\xi_{\text{CB}}^{\text{lim}} \equiv \xi_{\text{CB}}^{\rho}
\end{equation}
with
\begin{equation}
\xi_{\text{CB}}^{\rho} = -\log  \min_{0\leq s\leq1}\text{Tr}\left[(\rho^{(n)}_A)^s(\rho_E^{(n)})^{1-s}\right],
\end{equation}
 thereby allowing a direct evaluation in terms of the reduced density matrices
$\rho^{(n)}_{\text{E}}$ and $\rho^{(n)}_{\text{A}}$.
Using either of these 
 expressions for the QCB we can then identify correlator classes 
 in the thermodynamic limit: small values of $\xi_{\text{CB}}$ correspond to  large values of $I_n$ and
indicate a successful ansatz.


\section{Calculating $I$ using matrix-product states \label{sct:mps}}

\begin{figure}[t] 
   \centering
   \includegraphics[width=3in]{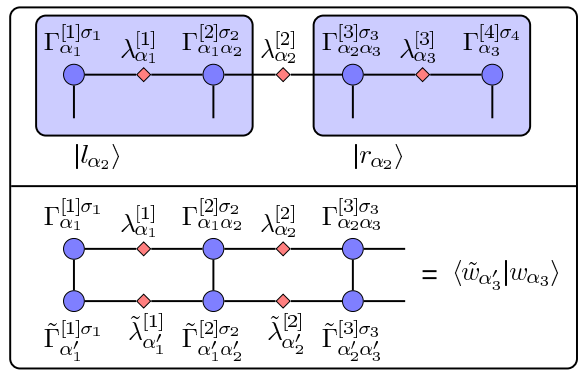} 
  \caption{Top: Schematic of the matrix product decomposition of a four-site lattice.  The circles indicate sites, $i$, with an applied tensor, $\Gamma^{[i]}$, and the diamonds denote bonds which carry Schmidt coefficients, $\lambda^{[i]}$, on the $i^{th}$ bond.   The shading indicates a decomposition 
  into left and right subsystems described by a renormalized basis $\vert l_{\alpha} \rangle$ and $\vert r_{\alpha} \rangle$ of eigenvectors of the respective reduced density matrices.  Bottom:  Schematic depicting an overlap of two matrix product states.  The top line corresponds to a state A (e.g., an ansatz state) in the matrix product representation while the bottom line corresponds to a second state E (e.g., an exact state).  }
  \label{fig1}
\end{figure}

Our simulation uses an MPS approximation to a 
state in the full spin Hilbert space.  The coefficients $c(\lbrace \sigma_i \rbrace)$ of the expansion of the state in the $\sigma^z$-basis,
\begin{equation}
|\Psi\rangle = \sum_{\lbrace \sigma_i \rbrace} c(\lbrace \sigma_i \rbrace) |\sigma_1\rangle \ldots  |\sigma_N\rangle
\end{equation}
are given as a product of matrices:
\begin{equation}
c(\lbrace \sigma_i \rbrace) = \sum_{\alpha_1, \ldots, \alpha_N} \Gamma_{\alpha_1}^{[1] \sigma_1} \lambda_{\alpha_1}^{[1]} \Gamma_{\alpha_1 \alpha_2}^{[2] \sigma_2} \lambda_{\alpha_2}^{[2]} \ldots \Gamma_{\alpha_N}^{[N] \sigma_N},
\end{equation}
where $\alpha$ indexes the auxiliary state space (of size $M$),  $\Gamma$ are rank three tensors that must be determined, and the coefficients $\lambda$ are the Schmidt eigenvalues of a bipartite splitting of the system at that site, i.e. they are equal to the eigenvalues of the reduced density matrices obtained by such a splitting. In the following we denote MPS states as $| \Psi \rangle = (\Gamma^{[i] \sigma_i}_{\alpha_i \beta_i}, \lambda^{[i]}_{\beta_i})$.

The accuracy of the MPS approximation depends on the decay of these eigenvalues and can be controlled by tuning the matrix dimension $M$. In the case of the Ising model in transverse field, the Schmidt coefficients are found to decay very quickly. We therefore perform our calculations with a matrix size $M=100$ and up to $N=64$ spins. Imaginary time evolution is used to project into the ground state. We apply a first-order Trotter decomposition with an initial time-step $d\tau = 0.05$, which is decreased to $d\tau=0.0001$ during the simulation. In what follows an "exact" state (E) refers to an MPS approximation to the exact ground state.  

Once the ground state has been found, we must obtain the density matrix in a common, orthonormal basis $\lbrace |v_a\rangle \rbrace$ for both the ansatz and exact states. We first join the two bases by concatenating 
Schmidt coefficients and the tensors blockwise, i.e. for two states $|\Psi \rangle = (\Gamma^{[i] \sigma_i}_{\alpha_i \beta_i}, \lambda^{[i]}_{\beta_i}), |\tilde{\Psi} \rangle = (\tilde{\Gamma}^{[i] \sigma_i}_{\alpha_i \beta_i}, \tilde{\lambda}^{[i]}_{\beta_i})$ we have $|\hat{\Psi} \rangle$ given by
\begin{eqnarray}
\hat{\Gamma}^{[i] \sigma}_{\alpha \beta} &=& \left\{
	\begin{array}{lr}
		\Gamma^{[i] \sigma}_{\alpha \beta} & \alpha, \beta \in \lbrace 1, \ldots, M \rbrace\\
		\tilde{\Gamma}^{[i] \sigma}_{\alpha \beta} & \alpha, \beta \in \lbrace M+1, \ldots, 2M \rbrace
	\end{array}
	\right. \\
\hat{\lambda}^{[i]}_{\alpha} &=& \left\{
	\begin{array}{lr}
		\lambda^{[i]}_{\alpha} & \alpha \in \lbrace 1, \ldots, M \rbrace\\
		\tilde{\lambda}^{[i]}_{\alpha} & \alpha \in \lbrace M+1, \ldots, 2M \rbrace.
	\end{array}
	\right.
\end{eqnarray}
We define the overlap matrix of two sets of states $| w_m \rangle = ( \Gamma^{[i]\sigma_i}_{\alpha \beta}, \lambda^{[i]}_{\beta})$ and $| \tilde{w}_m \rangle = ( \tilde{\Gamma}^{[i]\sigma_i}_{\alpha \beta}, \tilde{\lambda}^{[i]}_{\beta})$ describing some part of the system (bottom panel, Fig.~\ref{fig1}), which are taken to be the Schmidt eigenvectors of a bipartite decomposition of the system:
\begin{equation}
\langle \tilde{w}_{\alpha_n'} | w_{\alpha_n} \rangle =
\sum \mathcal{F} \left(
	\overline{  \prod_i \tilde{\Gamma}^{[i] \sigma_i}_{\alpha'_i \alpha'_i} \tilde{\lambda^{[i]}}_{\alpha'_i}  }
	\prod_i \Gamma^{[i] \sigma_i}_{\alpha_i \alpha_i} \lambda^{[i]}_{\alpha_i}  
 \right),
\end{equation}
where the summation runs over all orthogonal spin configurations and $\mathcal{F}$ indicates the summation over all remaining indices. This allows us to find a transformation that we can use to orthonormalize the basis of $|\hat{\Psi}\rangle$ for a specific bipartition.

The reduced density matrices can now be computed using
\begin{equation}
\rho^{\text{red}}_{a,b}=\sum_{\alpha,\beta,t} \lambda_{\alpha} \lambda_{\beta}\langle v_a \vert l_{\alpha} \rangle\langle l_{\beta} \vert v_{b} \rangle 
\langle r_t \vert r_{\alpha} \rangle \langle r_{\beta} \vert r_{t}\rangle.
\end{equation}
$|r_{\alpha}\rangle$  and $|l_{\alpha}\rangle$ denote states obtained from 
a right and left partitioning of the lattice (top panel, Fig.~\ref{fig1}). The sum over states $|r_t\rangle$ traces out the right
$N-n$ sites.  


\section{Ising chain in transverse field \label{sct:ising}}

\begin{figure}[t] 
   \centering
  \includegraphics[width=\columnwidth]{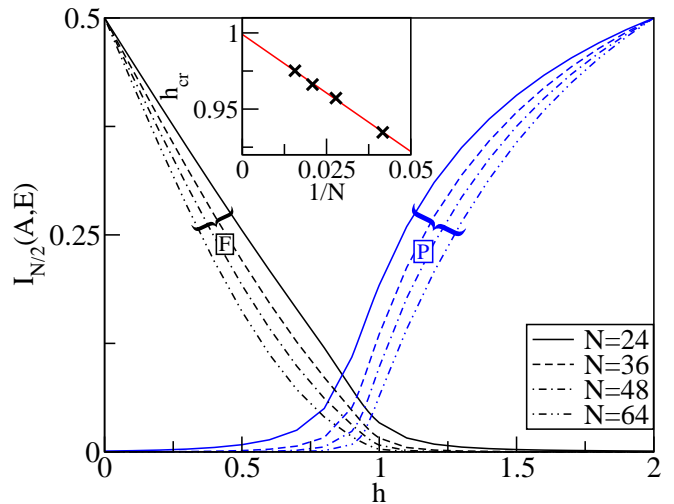} 
 \caption{(color online) Indistinguishability $I_{N/2}(A,E)$ plotted as a function of magnetic field $h$ for the one-dimensional transverse quantum Ising model for several different system sizes, $N$, and two different ansatz states, $A=F$, (ferromagnetic, $h_{\text{ref}}=0$, black lines) and $A=P$ (paramagnetic, 
 $h_{\text{ref}}=2$, blue lines).  The inset plots the crossing point of $I_{N/2}(F,E)$ and $I_{N/2}(P,E)$, with $\Psi_{\text{P}}$ evaluated for $h_{\text{ref}}=100$, versus $N^{-1}$.  A straight line fit yields a quantum critical point at $h_{\text{cr}}= 0.999(1)$ as $N\rightarrow\infty$, in agreement with standard results~\cite{sachdev1999}.  
}
\label{fig2}
\end{figure}

We first apply the indistinguishability measure to the simplest model with a quantum phase transition, the ferromagnetic transverse Ising model:
 \begin{eqnarray}
  H_{\text{Is}}=- \sum_{i}\sigma^x_{i}\sigma^x_{i+1}-h \sum_{i} \sigma^z_{i}. 
 \label{H}
\end{eqnarray}
Here $\sigma^{\alpha}_{i}, \alpha=x,y,z$ are the Pauli matrices and the sites $i$ are located on an $N$-site chain with open boundary conditions.  For a review of the properties of this model see Ref.~(\onlinecite{sachdev1999}) and references therein.  
  
Physically motivated ansatz wavefunctions can be defined for Eq.~(\ref{H}) by noting that
for $h>1$ the ground state is a paramagnet with exponentially decaying correlators, 
$\langle \sigma^x_{0}\sigma^x_{j} \rangle\sim \exp{(-j/\xi)}$, while for $h<1$ the ground state 
is in the ferromagnetically ordered phase with
long range  order,  $\langle \sigma^x_{0}\sigma^x_{j} \rangle\sim m^2$ for
$j\rightarrow \infty$, where $m$ is the spontaneous magnetization.  On the ferromagnetic side the exact $h\rightarrow 0$ ground state (one of the two degenerate ground states) is given by:
$\Psi_{\text{F}}=\prod_{i}\vert\uparrow\rangle_i$,
while on the paramagnetic side in the limit $h \rightarrow \infty$ we have:
$\Psi_{\text{P}}=\prod_{i}\vert\rightarrow\rangle_i$. These states are ansatz wavefunctions that we will apply at all values of magnetic field $h$. In the thermodynamic limit there is a quantum phase transition at $h=1$.  Without relying on the explicit behavior of any correlation functions, we will show using $I_{n}$ that for $h \neq 1$ our ansatz wavefunctions fall into two distinct correlator classes that characterize the two phases on either side of the transition.  $I_{n}$ thus allows an efficient test of the accuracy of ansatz states in reproducing all 
$n$-particle correlation functions of the exact state, without explicit calculation of these.
We further show that the location of the transition can be accurately identified.

We focus here on calculations for large values of $n$ that will allow us to analyze the scaling of $I_{n}$ when this is an intensive quantity.  Thus we consider $n = N/2$, where the total number of spins $N$ varies.  The indistinguishability $I_{n=N/2}$ of the exact ground state of Eq.~(\ref{H})
with the ferromagnetic and paramagnetic
ansatz states $\Psi_{\text{F}}$ and $\Psi_{\text{P}}$ was computed.  We use 
$\Psi_{\text{F}}=\Psi_{\text{E}}^{h_{\text{ref}}=0}$  for the ferromagnetic ansatz.  For most calculations it is sufficient to use
$\Psi_{\text{P}}=\Psi_{\text{E}}^{h_{\text{ref}}}$ with
${h_{\text{ref}}=2}$ for the paramagnetic ansatz but larger values of $h_{\text{ref}}$ will be used to extract information about the phase transition.  

The calculated indistinguishabilities $I_{N/2}(F,E)$ and $I_{N/2}(P,E)$ are shown in the main panel of Fig.~\ref{fig2} as a function of $h$ for several system sizes.    
For $h\lesssim1$, we find  $I_{N/2}(F,E)$ large with a weak decay with $N$.  In contrast, we find here   
a strong suppression of $I_{N/2}(P,E)$ as $N\rightarrow\infty$, implying that
an optimally chosen correlator
of up to $N/2$ particles will not successfully distinguish the exact state from 
$\Psi_{\text{A}}^{\text{F}}$ but will successfully distinguish the exact state from the paramagnetic state for large enough $N$. 
For $h\gtrsim1$, we find the reverse situation.

We can use $I_n$ to accurately identify the phase transition point, $h_\text{cr}$.  We search for the critical 
point by finding the $h$ at which $I_{N/2}(F,E)=I_{N/2}(P,E)$ and extrapolating to the 
thermodynamic limit.  The inset of Fig.~\ref{fig2} shows a linear extrapolation in $1/N$ that agrees 
with the known solution, $h_\text{cr}=1$.


\begin{figure}[t] 
   \centering
  \includegraphics[width=\columnwidth]{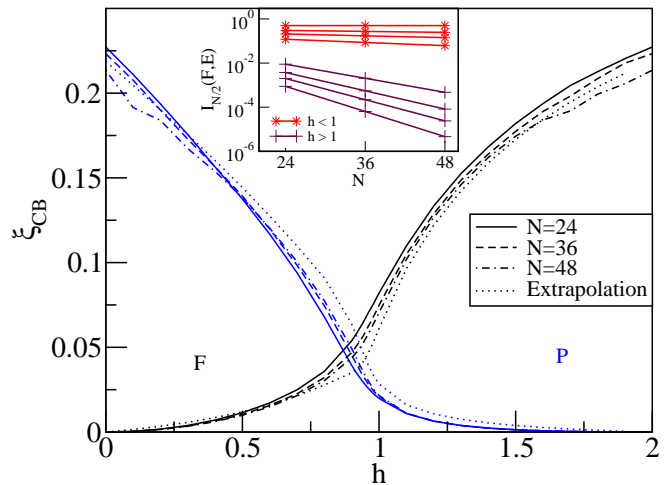}  
 \caption{(color online) The quantum Chernoff bound versus magnetic field using the extrapolation of indistinguishability, $\xi_{\text{CB}}^{\text{lim}}$, (dotted line) and the reduced density matrices directly,  $\xi_{\text{CB}}^{\rho}$, (solid, dashed and dot-dashed).  
 A suppression of $\xi_{\text{CB}}$ indicates success for the ferromagnetic, F, black lines (paramagnetic, P blue lines) ansatz for 
 $h\lesssim 1$ ($h\gtrsim 1$).   The inset plots shows a log plot of $I_{N/2}(F,E)$ versus $N$ for several $h$ to show an abrupt change in scaling near $h=1$.
 }
\label{fig3}
\end{figure}

We compute the QCB for each of the two phases in the transverse Ising model to demonstrate that the existence of two distinct correlator classes can also be found via the scaling exponent of $I_n$.  Fig.~\ref{fig3} 
plots the QCB versus $h$ evaluated with two different methods.  The dotted lines plot the finite size extrapolation of $\xi_{\text{CB}}^{\text{lim}}$ for both A=F ($h_{\text{ref}}=0$, black) and A=P ($h_{\text{ref}}=2$, blue).  The remaining lines show how the data collapse towards this line for several discrete $N$ values.  
We see that the scaling exponent, $\xi_{\text{CB}}$, remains finite in the 
thermodynamic limit and correctly identifies correlator classes on either side of the critical point.  
Precise location of the critical point from the QCB is complicated by the need to extrapolate an exponent and the associated numerical error.  Location from the scaling of $I_{n}$ as in Fig.~\ref{fig2} is more direct and appears more robust in this case.

The critical point can be defined in terms of the indistinguishability as the unique point
in parameter space that, for a given ansatz wavefunction, separates regions characterized by dramatically 
different scaling of $I_n$.  
As demonstrated above, both the direct evaluation of $I_n$ and evaluation of the QCB allow the critical point between two phases to be located as the point where the indistinguishability measures for the two different ansatz functions are equal.  To further characterize the critical point for finite sized systems we can 
define an indistinguishability susceptibility 
for $I_{n}(A,E)$ by
\begin{eqnarray} \label{eqn:susc}
\chi_{I_n} &=& \lim_{h \rightarrow ^\pm h_{\text{ref}}} d I_{n} (\Psi_{\text{E}}^{h_{\text{ref}}},\Psi_{\text{E}}^h)/dh \nonumber \\
\ &=& \lim_{\varepsilon \rightarrow 0} \frac{0.5 - I_n (\Psi_{\text{E}}^{h_{\text{ref}}}, \Psi_{\text{E}}^{h_{\text{ref}}\pm\varepsilon})} {\pm \varepsilon }
\end{eqnarray}
where we have used $\Psi_{\text A}=\Psi_{\text{E}}^{h_{\text ref}}$. Eq.~(\ref{eqn:susc}) 
should coincide with the maximum of the derivative of $I_{n}(A,E)$ for a given $h_{\text{ref}}$, when $\Psi_{\text A}=\Psi_{\text{E}}^{h_{\text ref}}$.
Our direct calculations of $I_{n}$ show that for the transverse Ising model, the critical point 
can be identified with a 
peak in $\chi_{I_{N/2}}$ versus $h$.  


\section{Spin-1 bilinear-biquadratic chain \label{sct:blbq}}

\begin{figure*}[t] 
   \centering
  \includegraphics[width=7in]{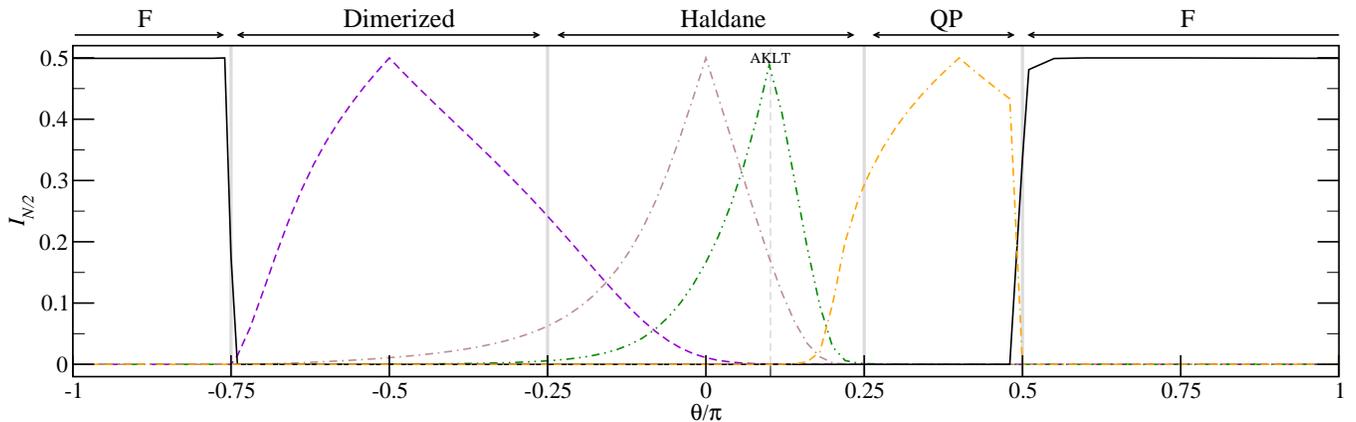} 
 \caption{Indistinguishability versus $\theta$ for the bilinear-biquadratic Heisenberg model, Eq.~\ref{Hbi}, with $N=36$ and $n=18$.  Distinct correlator classes surround 
 5 different reference states for which $I_{N/2}=1/2$: the ferromagnetic (solid), quadrupolar (dot-double dashed), Haldane (dot-dashed) and dimerized (dashed) phases.  The AKLT point (double dot-dashed) at $\theta = \tan^{-1}(1/3)$ appears within the Haldane phase. }
  \label{fig4}
\end{figure*}

We now apply our distinguishability measure to analyze a richer model with ground states 
characterized by more complicated correlators, the bilinear-biquadratic Heisenberg chain, 
defined by
\begin{eqnarray}
  H_{\text{bl}-\text{bq}}= \sum_{i} \left[ \cos{\theta}({\bf S}_{i}{\bf S}_{i+1})+ \sin{\theta}({ \bf S}_{i}{\bf S}_{i+1})^2\right], 
 \label{Hbi}
\end{eqnarray}
where ${\bf S}$ is the spin-$1$ operator 
and $\theta$ a parameter.  A growing body of analytic and numerical work has shown that this model hosts a variety of ground state phases (for a review see Ref.~(\onlinecite{laeuchli2006}) and references therein).
 
An integrable point \cite{affleck1987,affleck1988} at $\theta_{\text{AKLT}}=\tan^{-1}{1/3}$ has a particularly simple form for the exact ground state that belongs to a class of VBS wavefunctions related to the Laughlin ansatz state \cite{arovas1988}.  The VBS state at $\theta_{\text{AKLT}}$ can be written as
$\Psi_{\text{VBS}}=\prod_{i}(a^{\dagger}_i b^{\dagger}_{i+1}-b^{\dagger}_i a^{\dagger}_{i+1})\vert 0 \rangle$, where 
$a$ and $b$ annihilate Schwinger bosons defined by $S^x+iS^y=a^{\dagger}b, S^z=(a^{\dagger} a-b^{\dagger}b)/2,$ and $ a^{\dagger} a+b^{\dagger}b=2$.  $\Psi_{\text{VBS}}$ characterizes a state with exponentially decaying local correlators. The AKLT state is also characterized by hidden, long-ranged chain correlators \cite{girvin1989}. Notably, this state does not break translational symmetry. On a finite chain with open boundary conditions, a four-fold degeneracy appears which is related to open spin-$1/2$ degrees of freedom at the ends of the chain.

\begin{figure}[t]
	\centering
	   \vspace{0.5cm}
	\includegraphics[width=\columnwidth]{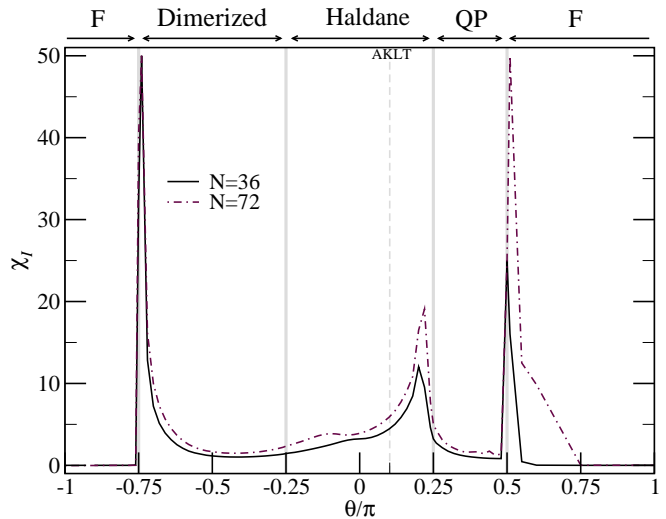}
	\caption{Indistinguishability susceptibility, Eq.~\ref{eqn:susc}, computed for the bilinear-biquadratic chain, Eq.~\ref{Hbi}.  Here we use $N=36,72$, $n = N/2$ and $\varepsilon = 0.02$ for the relevant part of the phase diagram. The peaks indicate phase transitions. In the case of discontinuous transitions, peaks remain finite only due to the discretization of the values of $h_{\text{ref}}$. \label{fig5} }
\end{figure}

We address the phases of this model from the point of view of ansatz states by taking
 five specific values of $\theta$ as reference points to capture the various possible phases. In particular, we choose $\theta_{\text{ref}}=\pi$ for the ferromagnetic (F) phase, $\theta_{\text{ref}}=0.4\pi$ for the quadrupolar (QP) phase, $\theta_{\text{ref}}=0$ and $\theta_{\text{ref}}=\theta_{\text{AKLT}}$ for the Haldane phase (corresponding to the Heisenberg and AKLT states) and $\theta_{\text{ref}}=-\pi/2$ for the dimerized phase.  In addition to these ansatzes defined by the ground states of the Hamiltonian Eq.~(\ref{Hbi}) for the five reference $\theta$ values, we shall also consider a trial wave function for a fully dimerized state, obtained at $\theta = -\pi/2$ with a modified Hamiltonian that results from omitting all even-bond terms in Eq.~(\ref{Hbi}).

Since, with the exception of the Haldane phase, analytic forms for the ground state wave function at these reference points are not known, the ansatz wavefunctions are given here by numerical solution for the ground state of Eq.~(\ref{Hbi}) at the reference values of $\theta_{\text{ref}}$. These numerical solutions $\Psi_A$ are generated with the matrix product approach of Section~\ref{sct:mps} at $\theta_{\text{ref}}$, just as the exact solutions $\Psi_E$ are generated at arbitrary values of $\theta$. This illustrates an important practical feature of the method, namely that we are not restricted to use of analytic ansatz functions.  We then calculate the indistinguishability measure, $I_{N/2}$, for system sizes $N=24\ldots72$ with open boundary conditions. 
Due to this choice of boundary conditions, we need to 
take $N$ as a multiple of 3 in order to be able to capture correlations at $k=2\pi/3$ which are important in the quadrupolar phase.

A typical result, for $N=36$, is presented in Fig.~\ref{fig4}, which shows 
$I_{N/2}$ as a function of $\theta$ for the five different correlator classes defined by the above values of $\theta_{\text{ref}}$.  The general variation of $I_{N/2}$ for each correlator class is consistent with what little is known about the phase boundaries in this system~\cite{laeuchli2006,girvin1989,solyom1987}, 
but also reveals new insights.
In particular,  we find several remarkable features.
First, the ferromagnetic ansatz is seen to be indistinguishable from the exact ground state over a wide range of $\theta$ values, $\theta \leq -3\pi/4$ and $\theta \geq +\pi/2$, with sharp, possibly first order, transitions signalling vanishing of the ferromagnetic state at $\theta=\pi/2$ and $-3\pi/4$.
 Second, the ground state of the Heisenberg point at $\theta=0$ is in the same correlator class as the AKLT state, 
supporting suggestions that there is a finite range of $\theta$ over which the ground state has the symmetry of the AKLT state~\cite{girvin1989}. 
Third, although the indistinguishability for $h_{\text{ref}}$ in the Haldane phase drops quickly as the dimerized phase is approached ($\theta \rightarrow -\pi/4$), the signature of the phase transition does not appear as strongly as the 
alternative Haldane to quadrupolar transition ($\theta \rightarrow +\pi/4$) for this system size.

Additionally, we have calculated the $n=N/2$ indistinguishability susceptibility, Eq.~\ref{eqn:susc}, for this 
model with $N=36$ and $N=72$ (Fig.~\ref{fig5}). The sharp transitions from the ferromagnetic state are 
reflected in large peaks in the susceptibility; the height of these peaks is controlled only by the discretization of the $\theta$ values.  These sharp peaks at $\theta=\pi/2$ and $-3\pi/4$  are consistent with the possibility of first order transitions out of the ferromagnetic phase. A well-pronounced transition also appears between the Haldane and the quadrupolar phase, 
with a pronounced shift due to finite size effects.  The indistinguishability susceptibility is thus sensitive to the difference between first order and continuous phase transitions, with the latter showing finite size shifts due to the divergence of the intrinsic length scales.

\begin{figure}[t]
	\centering
	   \vspace{0.5cm}
	\includegraphics[width=\columnwidth]{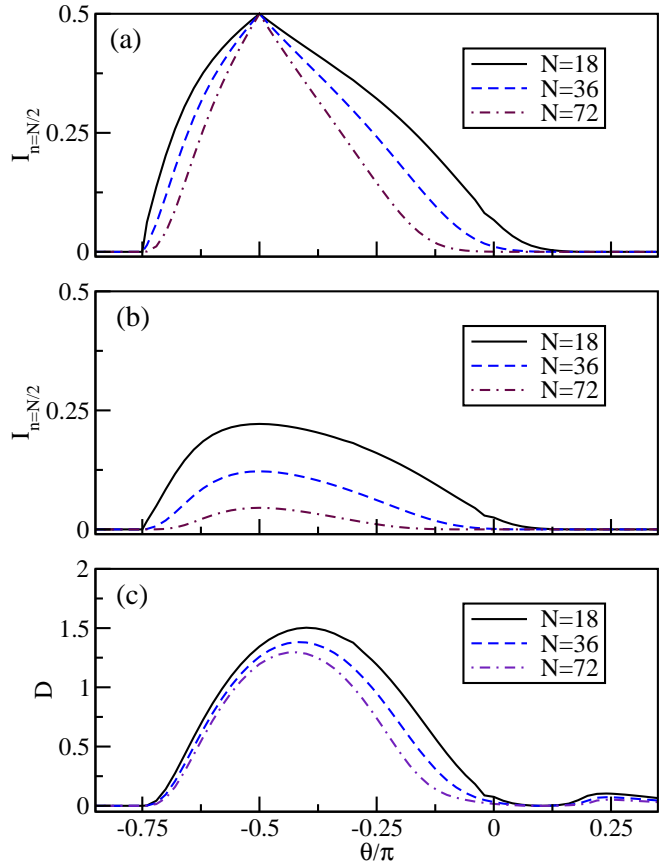}
	\caption{Analysis of the dimerized phase. (a) MPS ground state for $\theta = -\pi/2$ as the reference state. For small system sizes, the reference state remains a good ansatz state far beyond $\theta/\pi = -0.25$ into the Haldane regime. (b) $I_{N/2}$ for a strongly dimerized reference state obtained for $\theta = -\pi/2$ by omitting even-bond terms from the Hamiltonian, Eq.~(\ref{Hbi}). The indistinguishability remains small even around the maximally dimerized point near $\theta/\pi = -0.5$, indicating that a product of dimers only poorly characterizes the system. (c) Dimerization order parameter  (Eq.~(\ref{eqn:dimer})) for three system sizes. The dimerization remains finite far towards the AKLT point, rendering the Haldane and dimerized phases hard to distinguish on small length scales. \label{fig6}}
\end{figure}

We note that, in contrast to the clear signatures for transitions from the ferromagnetic phase and from the Haldane to quadrupolar phase, the transition from the dimerized phase to the Haldane phase is only very weak at these system sizes.  
A small peak does emerge at $N=72$, but for $N=36$ the peak corresponding to the transition appears considerably flattened out, almost to a plateau, and is also considerably shifted in location.  To understand this behaviour, we analyzed the fully dimerized ansatz state derived at $\theta = -\pi/2$ with the omission of all even-bond terms in Eq.~(\ref{Hbi}), as described above.
Fig. \ref{fig6} (b) shows that the indistinguishability for this ansatz state is relatively small, never exceeding 0.25, implying that this fully dimerized ansatz only poorly describes the dimerized phase of Eq.~(\ref{Hbi}), even in the proximity of the maximally dimerized point around $\theta/\pi = -0.5$. Our results thus confirm that there are strong fluctuations away from a simple state consisting entirely of products of dimers and that it is therefore difficult to precisely characterize the nature of the ground state in this parameter region.  

To analyze the degree of dimerization we define a dimerization order parameter:
\begin{equation} \label{eqn:dimer}
D = \frac{1}{N} \sum_{\text{bond}\ i} | H_i - H_{i+1} |
\end{equation}
with $H_i = \left[ \cos{\theta}({\bf S}_{i}{\bf S}_{i+1})+ \sin{\theta}({ \bf S}_{i}{\bf S}_{i+1})^2\right]$. In these units, the fully dimerized state is characterized by $D = 2.63$.  We can then demonstrate the degree of dimerization by direct evaluation of Eq.~\ref{eqn:dimer}.  This is plotted in Fig. \ref{fig6} (c), which shows a finite dimerization far into the Haldane regime.  For the smallest system sizes, the dimerization order parameter vanishes only at the AKLT point. This is due to the explicit breaking of translational symmetry induced by our use of open boundary conditions as well as to our restricted system sizes.     
We therefore expect that the weak peak in the susceptibility at $N=72$ should become more pronounced with larger system sizes or with periodic boundaries.

\section{Conclusion}

Motivated by the importance of ansatz wavefunctions for developing physical insight into strongly correlated phases where the numerically obtained wave function may be available but too complex for physical analysis, we have introduced an indistinguishability measure $I_n$ to assess the accuracy of ansatz functions by comparison with accurate numerical solutions.  The indistinguishability measure quantifies the ability of any set of $n$-particle correlators to distinguish two states, with a value $I_n>0$ 
in the thermodynamic limit showing that two states lie in the same correlator class and a value $I_n=0$ indicating that they lie in different classes.  The accuracy of an ansatz wavefunction can then be determined by evaluation of $I_n(A,E)$, the indistinguishability of the ansatz state $\Psi_A$ from the exact state $\Psi_E$. The real space scaling of the indistinguishability measure can be evaluated directly, or in terms of the quantum Chernoff bound \cite{audenaert2007}.
We demonstrated with two one-dimensional examples, the well-known transverse Ising chain and a spin-1 bilinear-biquadratic chain, that this allows 
physically motivated ansatz wavefunctions to be matched to accurate numerical wavefunctions in different phases.  We showed that the phase boundaries can be accurately obtained from the coincidence of $I_n(A,E)$ values for different ansatzes $\Psi_A$ and further defined an indistinguishability susceptibility that characterizes the location of the quantum phase transition.

The indistinguishability measure $I_{n}$ can be applied to analysis of ansatz wavefunctions using any of the many numerical techniques that 
are now available to efficiently obtain reduced density matrices, e.g., exact diagonalization, 
configuration interaction methods, 
density matrix renormalization group (DMRG) \cite{white,schollwoeck} and tensor network methods 
(e.g., the projected entangled-pair state \cite{verstraete2004}, multi-scale entanglement renormalization ansatz 
\cite{vidal2007}, etc.).   As an example of a direct application, our measure may 
be used with exact diagonalization \cite{morf1998} or 
DMRG \cite{shibata2001,feiguin2008} to study ansatz fractional quantum Hall states.  

We thank J. I. Korsbakken, A. L{\"a}uchli, S. Trebst and F. Verstraete for valuable discussions.  We thank the NSF PIF and SNSF for support.  Simulations were performed on the ETH Brutus cluster.  While this manuscript was under review, we became aware of subsequent work \cite{invernizzi2009} reporting related calculations for Ising chains.

\bibliography{bibliography}{}

\begin{thebibliography}{35}
\expandafter\ifx\csname natexlab\endcsname\relax\def\natexlab#1{#1}\fi
\expandafter\ifx\csname bibnamefont\endcsname\relax
  \def\bibnamefont#1{#1}\fi
\expandafter\ifx\csname bibfnamefont\endcsname\relax
  \def\bibfnamefont#1{#1}\fi
\expandafter\ifx\csname citenamefont\endcsname\relax
  \def\citenamefont#1{#1}\fi
\expandafter\ifx\csname url\endcsname\relax
  \def\url#1{\texttt{#1}}\fi
\expandafter\ifx\csname urlprefix\endcsname\relax\def\urlprefix{URL }\fi
\providecommand{\bibinfo}[2]{#2}
\providecommand{\eprint}[2][]{\url{#2}}

\bibitem[{\citenamefont{Wen}(1990)}]{wen1990}
\bibinfo{author}{\bibfnamefont{X.-G.} \bibnamefont{Wen}},
  \bibinfo{journal}{Int.\ J.\ Mod.\ Phys} \textbf{\bibinfo{volume}{B4}},
  \bibinfo{pages}{239} (\bibinfo{year}{1990}).

\bibitem[{\citenamefont{Haldane}(1983)}]{haldane1983}
\bibinfo{author}{\bibfnamefont{F.}~\bibnamefont{Haldane}},
  \bibinfo{journal}{Phys.\ Rev.\ Lett.} \textbf{\bibinfo{volume}{51}},
  \bibinfo{pages}{605} (\bibinfo{year}{1983}).

\bibitem[{\citenamefont{Trugman and Kivelson}(1985)}]{trugman1985}
\bibinfo{author}{\bibfnamefont{S.}~\bibnamefont{Trugman}} \bibnamefont{and}
  \bibinfo{author}{\bibfnamefont{S.}~\bibnamefont{Kivelson}},
  \bibinfo{journal}{Phys.\ Rev.\ B} \textbf{\bibinfo{volume}{31}},
  \bibinfo{pages}{5280} (\bibinfo{year}{1985}).

\bibitem[{\citenamefont{Jain}(2007)}]{jain2007}
\bibinfo{author}{\bibfnamefont{J.}~\bibnamefont{Jain}},
  \emph{\bibinfo{title}{Composite Fermions}} (\bibinfo{publisher}{Cambridge
  University Press}, \bibinfo{year}{2007}).

\bibitem[{\citenamefont{Affleck et~al.}(1987)}]{affleck1987}
\bibinfo{author}{\bibfnamefont{I.}~\bibnamefont{Affleck}} \bibnamefont{et~al.},
  \bibinfo{journal}{Phys.\ Rev.\ Lett.} \textbf{\bibinfo{volume}{59}},
  \bibinfo{pages}{799} (\bibinfo{year}{1987}).

\bibitem[{\citenamefont{Kitaev}(2003)}]{kitaev2003}
\bibinfo{author}{\bibfnamefont{A.}~\bibnamefont{Kitaev}},
  \bibinfo{journal}{Ann.\ Phys.} \textbf{\bibinfo{volume}{303}},
  \bibinfo{pages}{2} (\bibinfo{year}{2003}).

\bibitem[{\citenamefont{Laughlin}(1983)}]{laughlin1983}
\bibinfo{author}{\bibfnamefont{R.}~\bibnamefont{Laughlin}},
  \bibinfo{journal}{Phys.\ Rev.\ Lett.} \textbf{\bibinfo{volume}{50}},
  \bibinfo{pages}{1395} (\bibinfo{year}{1983}).

\bibitem[{\citenamefont{Morf}(1998)}]{morf1998}
\bibinfo{author}{\bibfnamefont{R.}~\bibnamefont{Morf}},
  \bibinfo{journal}{Phys.\ Rev.\ Lett.} \textbf{\bibinfo{volume}{80}},
  \bibinfo{pages}{1505} (\bibinfo{year}{1998}).

\bibitem[{\citenamefont{Helstrom}(1976)}]{helstrom1976}
\bibinfo{author}{\bibfnamefont{C.}~\bibnamefont{Helstrom}},
  \emph{\bibinfo{title}{Quantum Detection and Estimation Theory}}
  (\bibinfo{publisher}{Academic}, \bibinfo{year}{1976}).

\bibitem[{\citenamefont{Fuchs and van~de Graaf}(1999)}]{fuchs1999}
\bibinfo{author}{\bibfnamefont{C.}~\bibnamefont{Fuchs}} \bibnamefont{and}
  \bibinfo{author}{\bibfnamefont{J.}~\bibnamefont{van~de Graaf}},
  \bibinfo{journal}{IEEE\ Trans.\ Inf.\ Theory} \textbf{\bibinfo{volume}{45}},
  \bibinfo{pages}{1216} (\bibinfo{year}{1999}).

\bibitem[{\citenamefont{Korsbakken et~al.}(2007)}]{korsbakken2007}
\bibinfo{author}{\bibfnamefont{J.~I.} \bibnamefont{Korsbakken}}
  \bibnamefont{et~al.}, \bibinfo{journal}{Phys.\ Rev.\ A}
  \textbf{\bibinfo{volume}{75}}, \bibinfo{pages}{042106}
  (\bibinfo{year}{2007}).

\bibitem[{\citenamefont{Vidal}(2003)}]{vidal_TEBD}
\bibinfo{author}{\bibfnamefont{G.}~\bibnamefont{Vidal}},
  \bibinfo{journal}{Phys.\ Rev.\ Lett.} \textbf{\bibinfo{volume}{91}},
  \bibinfo{pages}{147902} (\bibinfo{year}{2003}).

\bibitem[{\citenamefont{Daley et~al.}(2004)\citenamefont{Daley, Kollath,
  Schollwock, and Vidal}}]{daley2004}
\bibinfo{author}{\bibfnamefont{A.~J.} \bibnamefont{Daley}},
  \bibinfo{author}{\bibfnamefont{C.}~\bibnamefont{Kollath}},
  \bibinfo{author}{\bibfnamefont{U.}~\bibnamefont{Schollwock}},
  \bibnamefont{and} \bibinfo{author}{\bibfnamefont{G.}~\bibnamefont{Vidal}},
  \bibinfo{journal}{Journal of Statistical Mechanics: Theory and Experiment}
  \textbf{\bibinfo{volume}{2004}}, \bibinfo{pages}{P04005}
  (\bibinfo{year}{2004}).

\bibitem[{\citenamefont{{\"O}stlund and Rommer}(1995)}]{ostromm}
\bibinfo{author}{\bibfnamefont{S.}~\bibnamefont{{\"O}stlund}} \bibnamefont{and}
  \bibinfo{author}{\bibfnamefont{S.}~\bibnamefont{Rommer}},
  \bibinfo{journal}{Phys.\ Rev.\ Lett.} \textbf{\bibinfo{volume}{75}},
  \bibinfo{pages}{3537} (\bibinfo{year}{1995}).

\bibitem[{\citenamefont{Verstraete et~al.}(2008)\citenamefont{Verstraete,
  Cirac, and Murg}}]{verstraete2008}
\bibinfo{author}{\bibfnamefont{F.}~\bibnamefont{Verstraete}},
  \bibinfo{author}{\bibfnamefont{J.~I.} \bibnamefont{Cirac}}, \bibnamefont{and}
  \bibinfo{author}{\bibfnamefont{V.}~\bibnamefont{Murg}},
  \bibinfo{journal}{Adv. Phys.} \textbf{\bibinfo{volume}{57}},
  \bibinfo{pages}{143} (\bibinfo{year}{2008}).

\bibitem[{\citenamefont{Furukawa et~al.}(2006)}]{furukawa2005}
\bibinfo{author}{\bibfnamefont{S.}~\bibnamefont{Furukawa}}
  \bibnamefont{et~al.}, \bibinfo{journal}{Phys.\ Rev.\ Lett.}
  \textbf{\bibinfo{volume}{96}}, \bibinfo{pages}{047211}
  (\bibinfo{year}{2006}).

\bibitem[{\citenamefont{Osborne and Nielsen}(2002)}]{osborne2002}
\bibinfo{author}{\bibfnamefont{T.}~\bibnamefont{Osborne}} \bibnamefont{and}
  \bibinfo{author}{\bibfnamefont{M.}~\bibnamefont{Nielsen}},
  \bibinfo{journal}{Phys.\ Rev.\ A} \textbf{\bibinfo{volume}{66}},
  \bibinfo{pages}{032110} (\bibinfo{year}{2002}).

\bibitem[{\citenamefont{Osterloh et~al.}(2002)}]{osterloh2002}
\bibinfo{author}{\bibfnamefont{A.}~\bibnamefont{Osterloh}}
  \bibnamefont{et~al.}, \bibinfo{journal}{Nature}
  \textbf{\bibinfo{volume}{416}}, \bibinfo{pages}{608} (\bibinfo{year}{2002}).

\bibitem[{\citenamefont{Vidal et~al.}(2003)}]{vidal2003}
\bibinfo{author}{\bibfnamefont{G.}~\bibnamefont{Vidal}} \bibnamefont{et~al.},
  \bibinfo{journal}{Phys.\ Rev.\ Lett.} \textbf{\bibinfo{volume}{90}},
  \bibinfo{pages}{227902} (\bibinfo{year}{2003}).

\bibitem[{\citenamefont{Zanardi and Paunkovic}(2006)}]{zanardi2006}
\bibinfo{author}{\bibfnamefont{P.}~\bibnamefont{Zanardi}} \bibnamefont{and}
  \bibinfo{author}{\bibfnamefont{N.}~\bibnamefont{Paunkovic}},
  \bibinfo{journal}{Phys.\ Rev.\ E} \textbf{\bibinfo{volume}{74}},
  \bibinfo{pages}{031123} (\bibinfo{year}{2006}).

\bibitem[{\citenamefont{Zhou and Barjaktarevic}(2008)}]{zhou2008}
\bibinfo{author}{\bibfnamefont{H.}~\bibnamefont{Zhou}} \bibnamefont{and}
  \bibinfo{author}{\bibfnamefont{J.}~\bibnamefont{Barjaktarevic}},
  \bibinfo{journal}{J.\ Phys.\ A} \textbf{\bibinfo{volume}{41}},
  \bibinfo{pages}{412001} (\bibinfo{year}{2008}).

\bibitem[{\citenamefont{Audenaert et~al.}(2007)}]{audenaert2007}
\bibinfo{author}{\bibfnamefont{K.}~\bibnamefont{Audenaert}}
  \bibnamefont{et~al.}, \bibinfo{journal}{Phys.\ Rev.\ Lett.}
  \textbf{\bibinfo{volume}{98}}, \bibinfo{pages}{160501}
  (\bibinfo{year}{2007}).

\bibitem[{\citenamefont{Sachdev}(1999)}]{sachdev1999}
\bibinfo{author}{\bibfnamefont{S.}~\bibnamefont{Sachdev}},
  \emph{\bibinfo{title}{Quantum Phase Transitions}}
  (\bibinfo{publisher}{Cambridge University Press,Cambridge, England},
  \bibinfo{year}{1999}).

\bibitem[{\citenamefont{Laeuchli et~al.}(2006)}]{laeuchli2006}
\bibinfo{author}{\bibfnamefont{A.}~\bibnamefont{Laeuchli}}
  \bibnamefont{et~al.}, \bibinfo{journal}{Phys.\ Rev.\ B}
  \textbf{\bibinfo{volume}{74}}, \bibinfo{pages}{144426}
  (\bibinfo{year}{2006}).

\bibitem[{\citenamefont{Affleck et~al.}(1988)\citenamefont{Affleck, Kennedy,
  Lieb, and Tasaki}}]{affleck1988}
\bibinfo{author}{\bibfnamefont{I.}~\bibnamefont{Affleck}},
  \bibinfo{author}{\bibfnamefont{T.}~\bibnamefont{Kennedy}},
  \bibinfo{author}{\bibfnamefont{E.}~\bibnamefont{Lieb}}, \bibnamefont{and}
  \bibinfo{author}{\bibfnamefont{H.}~\bibnamefont{Tasaki}},
  \bibinfo{journal}{Commun. Math. Phys.} \textbf{\bibinfo{volume}{115}},
  \bibinfo{pages}{477} (\bibinfo{year}{1988}).

\bibitem[{\citenamefont{Arovas et~al.}(1998)}]{arovas1988}
\bibinfo{author}{\bibfnamefont{D.}~\bibnamefont{Arovas}} \bibnamefont{et~al.},
  \bibinfo{journal}{Phys.\ Rev.\ Lett.} \textbf{\bibinfo{volume}{60}},
  \bibinfo{pages}{531} (\bibinfo{year}{1998}).

\bibitem[{\citenamefont{Girvin and Arovas}(1989)}]{girvin1989}
\bibinfo{author}{\bibfnamefont{S.}~\bibnamefont{Girvin}} \bibnamefont{and}
  \bibinfo{author}{\bibfnamefont{D.}~\bibnamefont{Arovas}},
  \bibinfo{journal}{Phys. Scr.} \textbf{\bibinfo{volume}{T27}},
  \bibinfo{pages}{156} (\bibinfo{year}{1989}).

\bibitem[{\citenamefont{S\'olyom}(1987)}]{solyom1987}
\bibinfo{author}{\bibfnamefont{J.}~\bibnamefont{S\'olyom}},
  \bibinfo{journal}{Phys. Rev. B} \textbf{\bibinfo{volume}{36}},
  \bibinfo{pages}{8642} (\bibinfo{year}{1987}).

\bibitem[{\citenamefont{White}(1992)}]{white}
\bibinfo{author}{\bibfnamefont{S.}~\bibnamefont{White}},
  \bibinfo{journal}{Phys.\ Rev.\ Lett.} \textbf{\bibinfo{volume}{69}},
  \bibinfo{pages}{2863} (\bibinfo{year}{1992}).

\bibitem[{\citenamefont{Schollw{\"o}ck}(2005)}]{schollwoeck}
\bibinfo{author}{\bibfnamefont{U.}~\bibnamefont{Schollw{\"o}ck}},
  \bibinfo{journal}{Rev.\ Mod.\ Phys.} \textbf{\bibinfo{volume}{77}},
  \bibinfo{pages}{259} (\bibinfo{year}{2005}).

\bibitem[{\citenamefont{Verstraete and Cirac}()}]{verstraete2004}
\bibinfo{author}{\bibfnamefont{F.}~\bibnamefont{Verstraete}} \bibnamefont{and}
  \bibinfo{author}{\bibfnamefont{J.~I.} \bibnamefont{Cirac}},
  \bibinfo{note}{cond-mat/0407066}.

\bibitem[{\citenamefont{Vidal}(2007)}]{vidal2007}
\bibinfo{author}{\bibfnamefont{G.}~\bibnamefont{Vidal}},
  \bibinfo{journal}{Phys. Rev. Lett.} \textbf{\bibinfo{volume}{99}},
  \bibinfo{pages}{220405} (\bibinfo{year}{2007}).

\bibitem[{\citenamefont{Shibata and Yoshioka}(2001)}]{shibata2001}
\bibinfo{author}{\bibfnamefont{N.}~\bibnamefont{Shibata}} \bibnamefont{and}
  \bibinfo{author}{\bibfnamefont{D.}~\bibnamefont{Yoshioka}},
  \bibinfo{journal}{Phys. Rev. Lett.} \textbf{\bibinfo{volume}{86}},
  \bibinfo{pages}{5755} (\bibinfo{year}{2001}).

\bibitem[{\citenamefont{Feiguin et~al.}(2008)}]{feiguin2008}
\bibinfo{author}{\bibfnamefont{A.}~\bibnamefont{Feiguin}} \bibnamefont{et~al.},
  \bibinfo{journal}{Phys.\ Rev.\ Lett.} \textbf{\bibinfo{volume}{100}},
  \bibinfo{pages}{166803} (\bibinfo{year}{2008}).

\bibitem[{\citenamefont{Invernizzi and Paris}(2009)}]{invernizzi2009}
\bibinfo{author}{\bibfnamefont{C.}~\bibnamefont{Invernizzi}} \bibnamefont{and}
  \bibinfo{author}{\bibfnamefont{M.}~\bibnamefont{Paris}},
  \bibinfo{journal}{arxiv:0905.0980}  (\bibinfo{year}{2009}).

\end{thebibliography}
\bibliographystyle{apsrev}

\end{document}